\documentclass[aps,pre,twocolumn,superscriptaddress,showpacs]{revtex4-2}
\usepackage{amssymb,amsmath,amsfonts,graphicx,epsf}
\usepackage{color}

\begin{document}

\title{Particle acceleration by magnetic Rayleigh-Taylor instability: mechanism for flares in black-hole accretion flows}

\author{Vladimir Zhdankin}
\email{vzhdankin@flatironinstitute.org}
\affiliation{Center for Computational Astrophysics, Flatiron Institute, 162 5th Avenue, New York, NY 10010, USA}

\author{Bart Ripperda}
\altaffiliation{NASA Hubble Fellowship Program, Einstein Fellow}
\affiliation{School of Natural Sciences, Institute for Advanced Study, 1 Einstein Drive, Princeton, NJ 08540, USA}
\affiliation{Center for Computational Astrophysics, Flatiron Institute, 162 5th Avenue, New York, NY 10010, USA}

\author{Alexander A. Philippov}
\affiliation{Department of Physics, University of Maryland, College Park, MD 20742, USA}

\date{\today}

\begin{abstract}
We study the magnetic Rayleigh-Taylor instability in relativistic collisionless plasma, as an astrophysical process for nonthermal particle acceleration. We consider dense plasma on top of a highly magnetized cavity with sheared magnetic field. Using particle-in-cell simulations, we show that small plumes grow and merge progressively to form a large-scale plume, which broadens to drive rapid magnetic reconnection in the cavity. We find that this leads to efficient particle acceleration capable of explaining flares from the inner accretion flow onto the black hole Sgr A*.
\end{abstract}


\maketitle

\section{Introduction}

{The Rayleigh-Taylor instability (RTI) is a fundamental macroscopic instability in neutral fluids and plasmas, broadly relevant to laboratory experiments, space physics, and astrophysics \citep{zhou_2017a, zhou_2017b, zhou_etal_2021}. It occurs at the interface of heavy matter that lies on top of light matter in a gravitational field (or accelerating frame), when other forces (from magnetic tension, rotation, shear flow, etc.) are insufficient to stabilize the configuration. The RTI develops fingers that grow into plumes and mix the matter. Ultimately, it will convert free gravitational energy into fluid energy.}

The RTI recently received attention as a process in accretion flows onto supermassive black holes such as Sgr A*, where it is a candidate mechanism for flares. Sgr A* shows daily bright and rapid flares in X-ray, near infrared (NIR), and submillimeter wavelengths \citep{yusef-zadeh_etal_2009, dodds-eden_etal_2009, ponti_etal_2017, boyce_etal_2022}. NIR flares were associated with a hot spot comparable to the size of the event horizon, orbiting near the black hole \citep{abuter_etal_2018}. 

The accretion flow of Sgr A* is conjectured to transiently {become a magnetically arrested disk (MAD)} in the inner region \citep[e.g.,][]{Ressler_2020, akiyama_etal_2022}. {In a MAD, the accumulation of magnetic flux near the black hole will regulate the accretion process by excavating the flow \citep{bisnovatyi-kogan_ruzmaikin_1974, narayan_etal_2003}, as confirmed by magnetohydrodynamic (MHD) simulations \citep{igumenshchev_etal_2003, tchekhovskoy_etal_2011, mckinney_etal_2012}.} Magnetic flux eruptions, seen in general relativistic magnetohydrodynamic (MHD) simulations of {MADs}, have been proposed to power flares \citep{dexter_etal_2020, porth_etal_2021}. More specifically, during a flux eruption, magnetic reconnection at the jet base (near the event horizon) can transform toroidal magnetic field into a large flux tube threaded by a strong vertical (poloidal) field that contains reconnection-energized relativistic plasma originating from the jet \citep{Ripperda2020}. An example of the flux tube formed by an eruption event in an MHD simulation is shown in Fig.~\ref{fig:visual}a, using data from Ref.~\citep{ripperda_etal_2022}; see Supplemental Material for animations of the simulations \footnote{See Supplemental Material at [URL will be inserted by publisher] for animations of the MHD and PIC simulations.}. This picture is further supported by NIR polarization measurements that imply a dominant vertical magnetic field component in the emitting region of the hot spots \citep{jimenez-rosales_etal_2020}. After a flux tube is ejected into the accretion flow, RTI may occur on the interface between the low-density, magnetically dominated flux tube and the denser accretion flow, as visualized in Fig.~\ref{fig:visual}a. To be a viable explanation for NIR flares, the RTI must efficiently dissipate free energy into relativistic nonthermal particles. 

Previous studies considered MHD simulations \citep[][]{wang_nepveu_1983, wang_robertson_1985, jun_etal_1995,bucciantini_etal_2004, stone_gardiner_2007, stone_gardiner_2007b, porth_etal_2014, carlyle_hillier_2017, skoutnev_etal_2021, briard_etal_2022, braileanu_etal_2023} or linear theory \citep[][]{lyubarsky_2010,ruderman_etal_2014, ruderman_2017, jiang_jiang_2019} of {the magnetic} RTI{, with applications to phenomena such as solar prominences \citep[e.g.,][]{isobe_etal_2005, keppens_etal_2015, hillier_2018, jenkins_keppens_2022} and inertial confinement fusion \citep[e.g.,][]{srinivasan_etal_2012, khiar_etal_2019, walsh_2022}.} However, accretion flows onto black holes such as Sgr A* are expected to be essentially collisionless, necessitating a kinetic model (i.e., the Vlasov-Maxwell equations) to {properly} describe the dissipation {and dynamics}. {Furthermore, relativistic effects may influence the RTI in this environment.}

To determine whether the RTI is a viable process for nonthermal particle acceleration in black-hole accretion flows (and other high-energy astrophysical systems such as pulsar wind nebulae and relativistic jets), this work investigates the RTI with local kinetic particle-in-cell (PIC) simulations. We choose physical parameters relevant for black-hole accretion flows: a magnetically sheared interface between a moderately sub-relativistic plasma (with comparable magnetic and plasma pressures) and a hot, magnetically dominated cavity. We show that the nonlinear development of the RTI in this physical regime leads to self-organization and large-scale magnetic reconnection that has the necessary ingredients to explain the NIR flares in Sgr A*.

\section{Methods}

We perform local electromagnetic PIC simulations of the RTI with the code {\em Zeltron} \citep{cerutti_etal_2013}. PIC simulations provide a self-consistent model of collisionless plasma dynamics \citep{birdsall_langdon_2004}. Here we provide an overview of the numerical setup.

We consider a collisionless plasma in a uniform gravitational field $\boldsymbol{g} = - g \hat{\boldsymbol{y}}$ in a square 2D Cartesian domain with coordinates $0 < x < L$ and $0 < y < L$ \footnote{We explored other aspect ratios, but found that growth of the RTI plumes is limited by the smaller dimension at late times, with the key results remaining similar}. Gravity is implemented as a particle force $\boldsymbol{F}_g=m \boldsymbol{g}$ added to the usual Lorentz force, where $m$ is the particle mass. To simulate this with periodic boundary conditions, we also evolve a reflected copy of the initial state with a reversed gravitational field $-\boldsymbol{g}$ in the region $-L < y < 0$; this reflected copy provides an independent realization of the RTI and enables the simulations to be performed with periodic boundary conditions (alternatively, one may use reflecting boundaries at $y=0$ and $y = L$). In this work, we only present the $y > 0$ domain. We include a buffer zone with $g = 0$ near the boundaries (for $|y| < L/4$ and $|y| > 3L/4$) to reduce interaction between the two domains.

The system is initialized with a ``cold'' slab of dense plasma (at $y > L/2$) on top of a ``hot'' cavity of dilute plasma (at $y < L/2$), such that $n_h/n_c \ll 1$ where $n_s$ are the uniform initial particle number densities in each region; henceforth the subscript $s \in \{ c, h \}$ will refer to ``cold'' and ``hot'' plasmas. The interface at $y = L/2$ has a sharp density discontinuity (at the cell scale). To satisfy stratified pressure equilibrium in the uniform density regions, we choose a linear temperature profile, $T(y) = T_{\rm int} - m g (y - L/2)$, in the regions where $g \neq 0$; elsewhere $T$ is uniform. Here, $T_{\rm int}$ is the (continuous) temperature at the interface.

The initial magnetic field is directed perpendicular to gravity, with uniform magnitudes $B_h$ and $B_c$ in the hot and cold regions, respectively, and a sharp discontinuity at $y = L/2$. The ratio of magnetic field magnitudes is determined by pressure balance at the interface, $B_h^2/B_c^2 = 1 + \left(1 - n_h/n_c \right) \beta_c$, where $\beta_c = 8 \pi n_c T_{\rm int} / B_c^2$ is the plasma beta of the cold region near the interface. There is a gradual rotation of the magnetic field direction by a shear angle $\theta_{\rm rot} = \cos^{-1}{(\boldsymbol{B}_h \cdot \boldsymbol{B}_c / B_h B_c)}$, across a distance $L/4$ centered on the interface; here, $\boldsymbol{B}_s = B_{s,x} \hat{\boldsymbol{x}} + B_{s,z} \hat{\boldsymbol{z}}$ denotes the magnetic field vector far from the interface. The magnetic field in the rotation region $(3/8)L < y < (5/8)L$ is calculated in two steps. In the first step, each magnetic field component is obtained from a linear interpolation between the bounding values of $\boldsymbol{B}_h$ (at $y = 3L/8$) and $\boldsymbol{B}_c$ (at $y = 5L/8$). In the second step, the magnitude of the magnetic field is rescaled so that it equals either $B_h$ (in the hot region) or $B_c$ (in the cold region). 

Since the simulations are 2D in space, there is an important degree of freedom in choosing the initial orientation of the domain with respect to the magnetic field vectors. In some orientations, the RTI may be artificially inhibited, if the unstable wavevectors are not within the 2D domain. Furthermore, only the in-plane magnetic field can be dissipated by magnetic reconnection; the out-of-plane magnetic flux is conserved due to the periodic boundary conditions. Thus, to allow maximal dissipation of magnetic flux through magnetic reconnection, we orient the domain such that in-plane fields are opposite, $B_{h,x} = -B_{c,x} \equiv B_{\rm in}$, where
\begin{align}
B_{\rm in} = \frac{B_h B_c \sin{\theta_{\rm rot}}}{(B_h^2 + B_c^2 + 2 B_h B_c \cos{\theta_{\rm rot}})^{1/2}} \, ,
\end{align}
while the out-of-plane components are given by $B_{s,z} = (B_s^2 - B_{\rm in}^2)^{1/2}$.

The RTI in our simulations is seeded by PIC noise, associated with the finite number of macroparticles. PIC noise has power over a broad range of wavenumbers in Fourier space. We confirmed that the results are similar when instead a large-scale perturbation is applied to the magnetic field at the interface.

We consider electron-positron (pair) plasma, as expected from pair production at the cavity source {\citep[e.g.,][]{moscibrodzka_etal_2011}}. We fix physical parameters $\beta_c = 4$, $g = {2.4} T_{\rm int} / m L$, $T_{\rm int} = m c^2/16$, and focus on a moderate shear angle of $\theta_{\rm rot} = \pi/4$; we point the reader to the appendix (Sec.~\ref{sec:rot}) for discussion on the effect of varying $\theta_{\rm rot}$. The value of $g$ above was chosen such that there is approximately one pressure scale height in the domain; if $g$ was chosen much larger, then the required temperature would become negative in the cold plasma, causing an inconsistency. We vary density ratio $n_c/n_h \in \{ 8, 31, 127, 511 \}$, with $n_c/n_h = 511$ the fiducial value. The magnetization parameter {$\sigma \equiv B^2/4 \pi n m c^2$ in the hot cavity} can be estimated as $\sigma_h \sim (5/32) n_c/n_h$ for these parameters, so the density ratios correspond to $\sigma_h \in \{ 1.3, 5, 20, 80 \}$ {(fiducial value $\sigma_h = 80$)}. Since $\sigma_h > 1$ for all cases, the cavity is magnetically dominated. We considered several different values of dimensionless system size, $L/\rho_c \in \{ 100 , 200, 400, 800 \}$, where $\rho_c = m v_T / e B_c$ is the initial (non-relativistic) characteristic Larmor radius in the cold slab and {$v_T = (3 T_{\rm int}/m)^{1/2}$} is the initial thermal velocity. The fiducial value is $L/\rho_c = 400$. All results are for the fiducial case unless otherwise noted.

\begin{figure} 
\includegraphics[width=0.95\columnwidth]{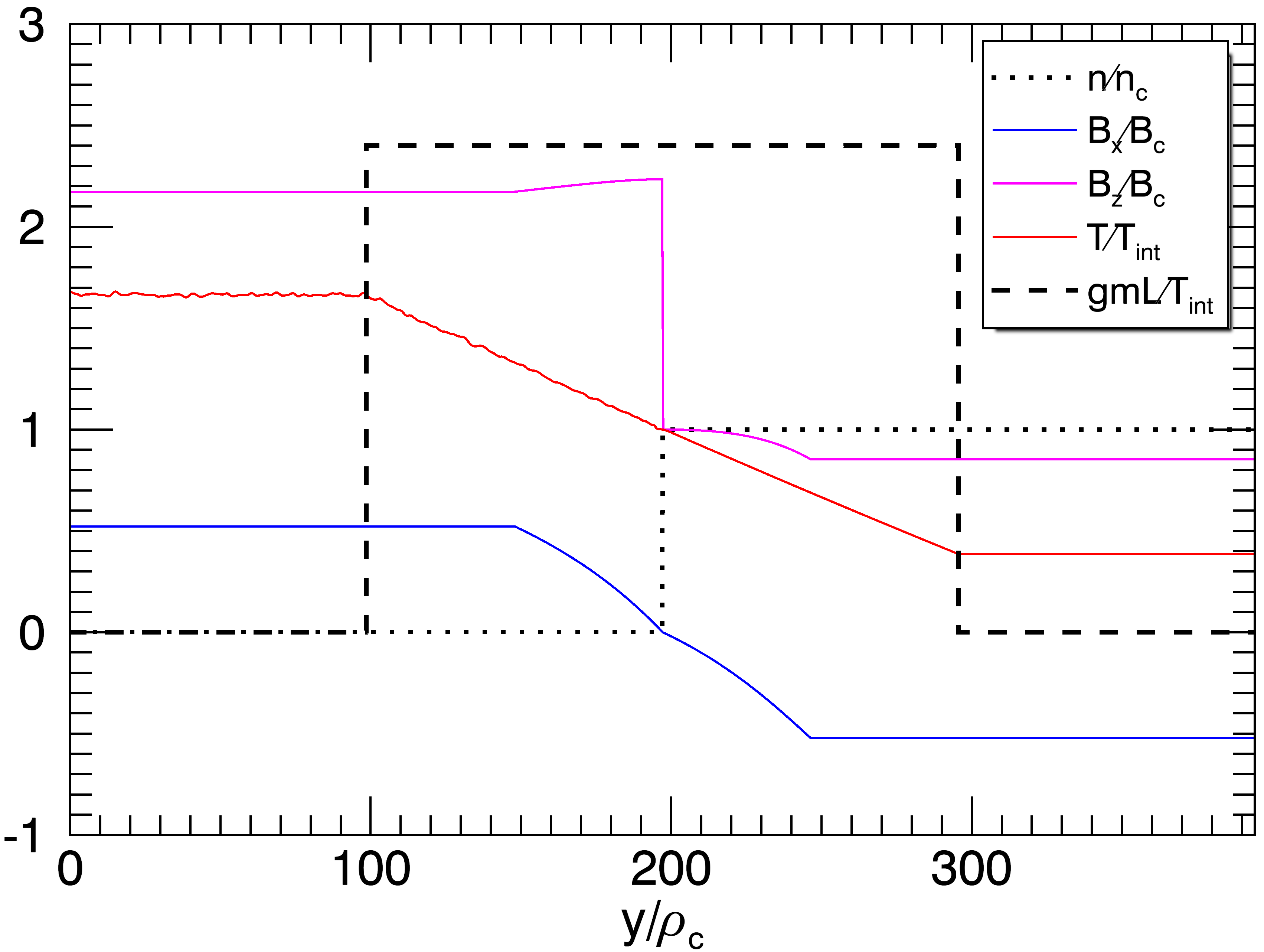}
   \centering
   \caption{\label{fig:profiles1d}
Initial field profiles of fiducial simulation in $y$ direction. The quantities are normalized particle number density $n/n_c$ (dotted), in-plane field $B_x/B_c$ (blue), out-of-plane field $B_z/B_c$ (magenta), temperature $T/T_{\rm int}$ (red), and gravitational field strength $g m L/T_{\rm int}$ (dashed). The gravitational force acts in the $-\hat{\boldsymbol{y}}$ direction. The $y < 0$ domain is obtained by reflecting these profiles across $y = 0$ and making $g$ negative.}
\end{figure}

In Fig.~\ref{fig:profiles1d}, we show the initial field profiles in the fiducial simulation, for the primary domain $0 < y < L$, where the cold dense slab ($L/2 < y < L$) lies of top of the hot cavity ($0 < y < L/2$). Note that since $n_c/n_h = 511$ in the fiducial simulation, $n_c$ appears as nearly zero in Fig.~\ref{fig:profiles1d}.

Numerical resolution is set by $\rho_c = 3 \sqrt{3} \Delta x$ where $\Delta x$ is the cell size. The simulated macroparticles all have equal weights (but we confirmed that results are similar when using low-weight particles in the cavity). Convergence studies indicate that 2 (electron+positron) particles per cell in the cavity is adequate. Thus, we choose 2 particles per cell in the cavity for $n_c/n_h \ge 31$, and 4 particles per cell for $n_c/n_h = 8$; the number of particles per cell in the cold slab is scaled in proportion to $n_c/n_h$.

It is important to acknowledge that our PIC simulations of RTI have several limitations, which we now list. 1) Since the model is local, the dynamics will generally be influenced by the boundaries at late times, due to either the growth of the plumes or the fall of material. Thus, the late-time evolution becomes artificial. However, we expect that the formation of a large-scale magnetized plume that must dissipate via magnetic reconnection to be a robust feature. 2) The gravitational field is assumed to be uniform (between the buffer regions) in our local setup, whereas it has a radial dependence in the global accretion problem. However, we anticipate that the development of the RTI occurs at small scales relative to the variation of the gravitational field, so the uniform field approximation is adequate until late times. 3) Effects of background rotation and shear are neglected for simplicity. Rotational support of the accretion flow may counteract gravity, and thus weaken the RTI. Shear will distort the plumes and possibly trigger the Kelvin-Helmholtz instability, which would compete with the RTI. These features may be studied in a shearing box framework \citep[e.g.,][]{bacchini_etal_2022}. 4) In the cold slab, an electron-proton plasma would be more appropriate than a pair plasma. For an ambient proton temperature $T_p \sim 0.1 m_p c^2$ in Sgr A*, where $m_p$ is the proton rest mass, the electrons and positrons would have a relativistic temperature, $T_e \sim T_p \sim (m_p/m_e) m_e c^2 \sim 200 m_e c^2$ (neglecting two-temperature effects). With this relativistic temperature, it would be easier to accelerate particles to the near-infrared emission range than in our current setup with $T/m_e c^2 = 1/16$. However, we anticipate that the macroscopic RTI dynamics are controlled by the sub-relativistic ions and thus should remain similar to our current cases. 5) The 2D domain may limit some dynamics (including turbulence) compared to 3D, as noted in the conclusions section.

\section{Results}

\begin{figure}
\includegraphics[width=\columnwidth]{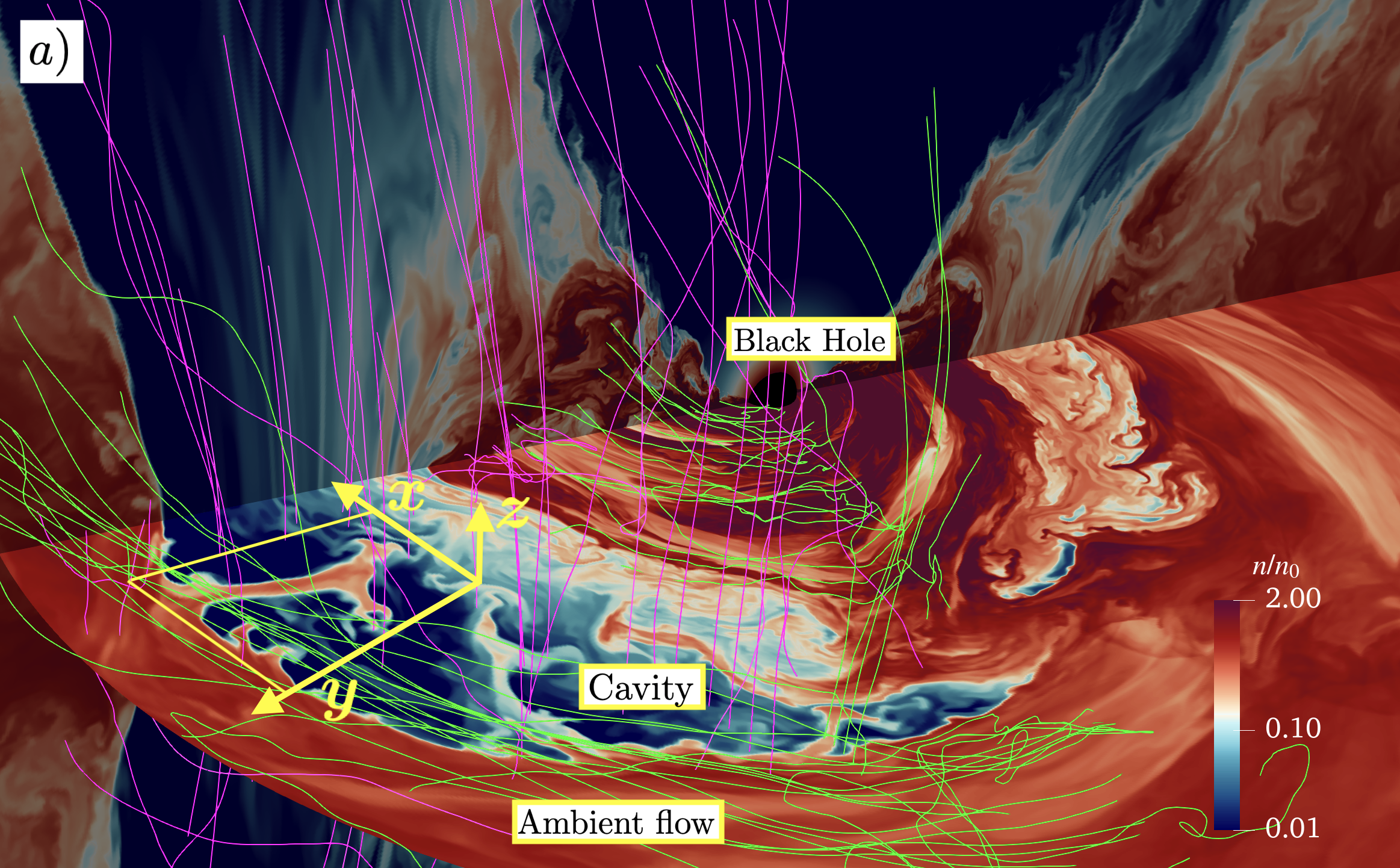}
\includegraphics[width=\columnwidth]{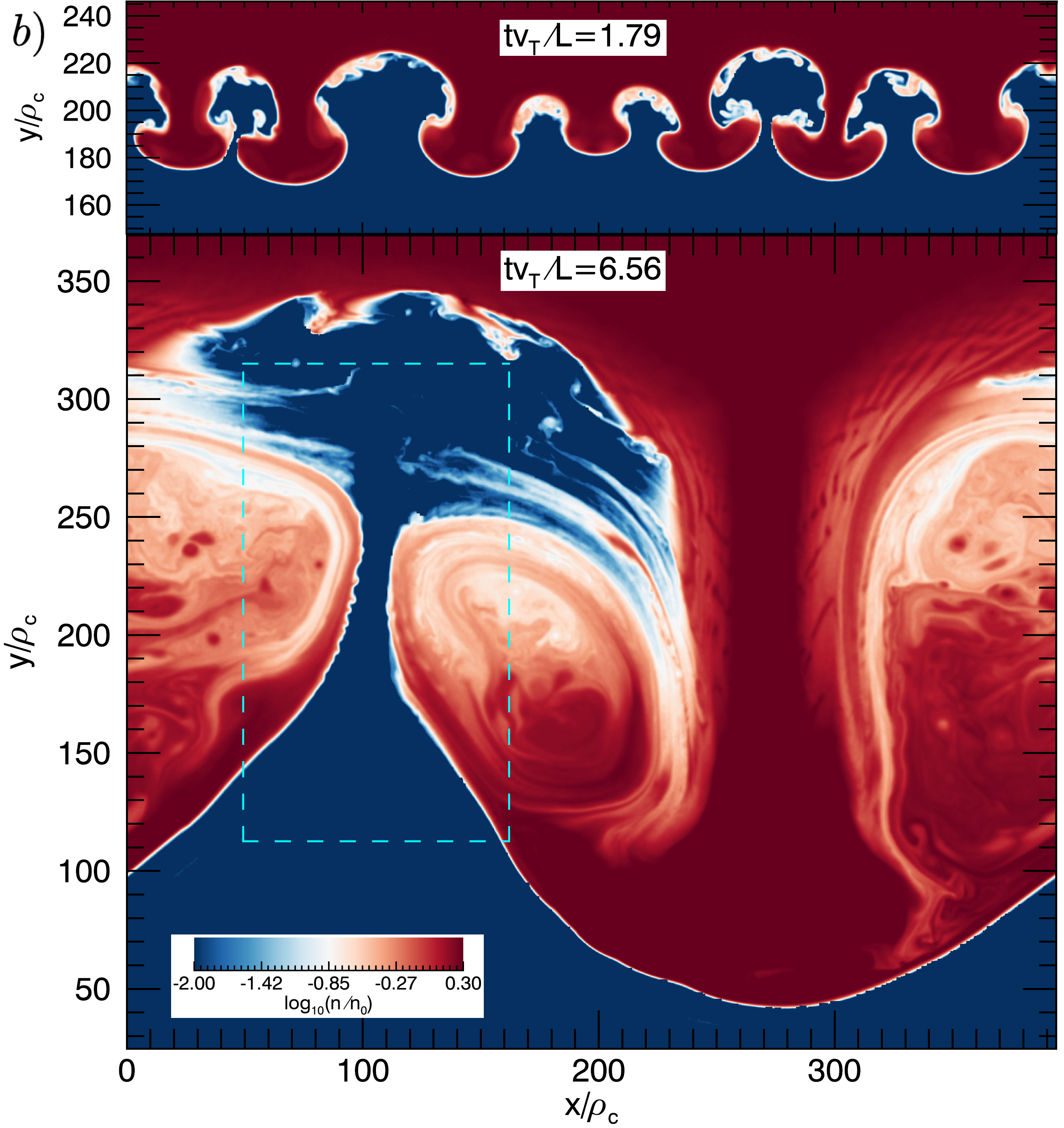}
   \centering
   \caption{\label{fig:visual} 
   (a) Density image showing accretion flow and cavity, with plumes on the RTI-unstable interface, in {cross-sections of a} global MHD simulation (using data from \citep{ripperda_etal_2022}; outer radial boundary is $33 r_g$ where $r_g$ is the gravitational radius). A sample of magnetic field lines in the ambient flow (green) and cavity (magenta) is overlaid. {The local coordinate system (neglecting tilt) is represented by the yellow axes.}
   (b) Density image at early times ($tv_T/L=1.8$) and late times ($tv_T/L=6.6$) {in the PIC simulation}; note that $y$ {extent} is cropped. The cyan box indicates the large-scale magnetic reconnection event shown in Fig.~\ref{fig:reconnection}.} 
\end{figure}

The simulations initially form small fingers at the interface, which evolve nonlinearly into plumes. These small plumes merge into progressively larger plumes until growing to the scale height, after which a single large plume remains (Fig.~\ref{fig:visual}b). This evolution is reminiscent of {that described in classical hydrodynamic \citep{young_etal_2001} and MHD studies of RTI} \citep{wang_nepveu_1983, wang_robertson_1985}. The merging process drives magnetic reconnection and turbulence, causing the final plume to have a complex morphology tainted by numerous secondary instabilities. In particular, {the interface develops asymmetric magnetic reconnection sites (reminiscent of Ref.~\citep{mbarek_etal_2022})}, shear flows (unstable to Kelvin-Helmholtz as in Ref.~\citep{sironi_etal_2021}), and beams (enabling kinetic pressure-anisotropy instabilities such as mirror and firehose). After it forms, the final plume broadens out, which leads to a rapid large-scale (tearing unstable) reconnection event at $t \approx 6.5 L/v_T$. {Although the plume extends into the $g = 0$ buffer zone, we confirmed that results are similar without the buffer zone.}

\begin{figure}
\includegraphics[width=\columnwidth]{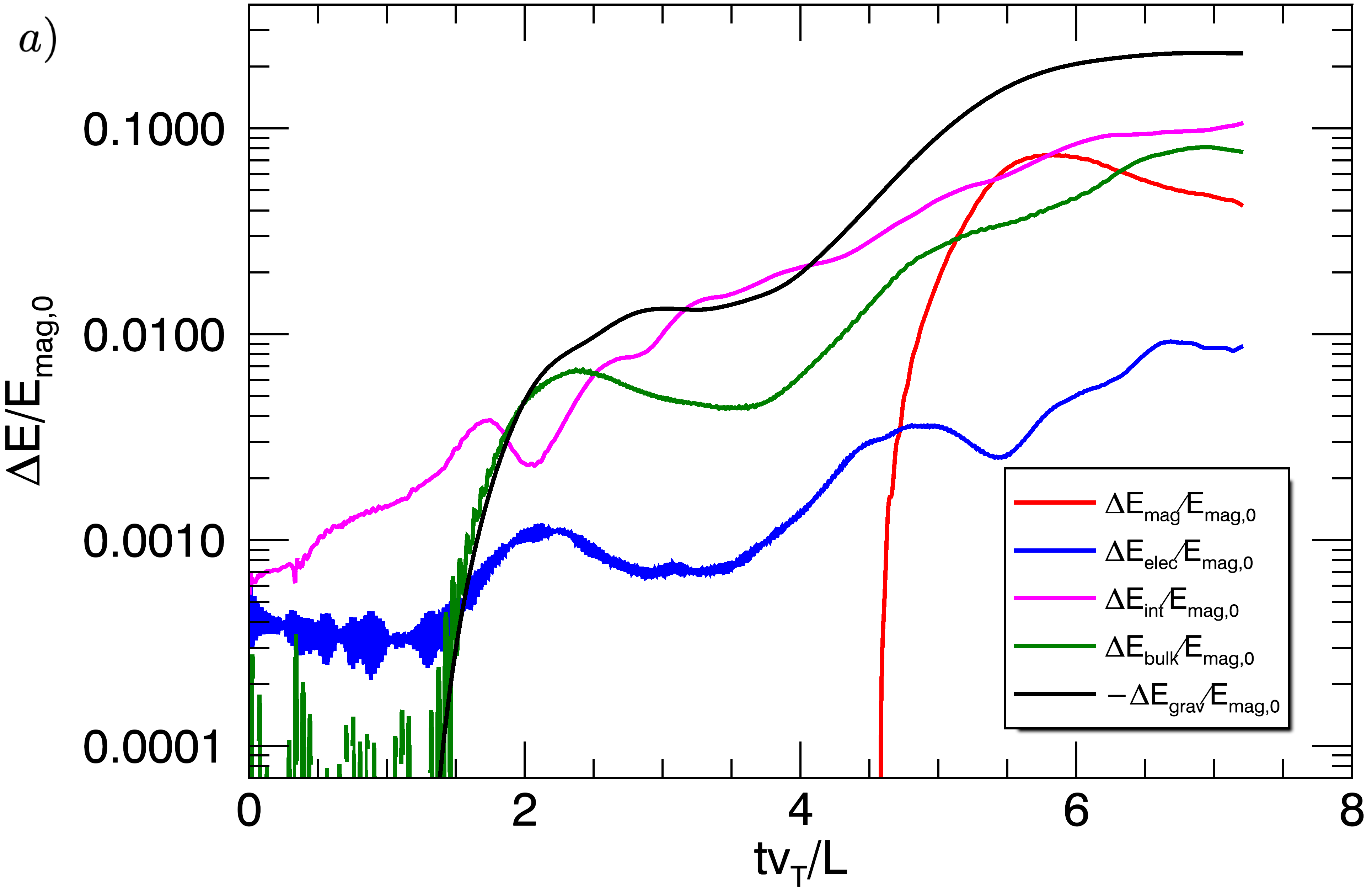}
\includegraphics[width=\columnwidth]{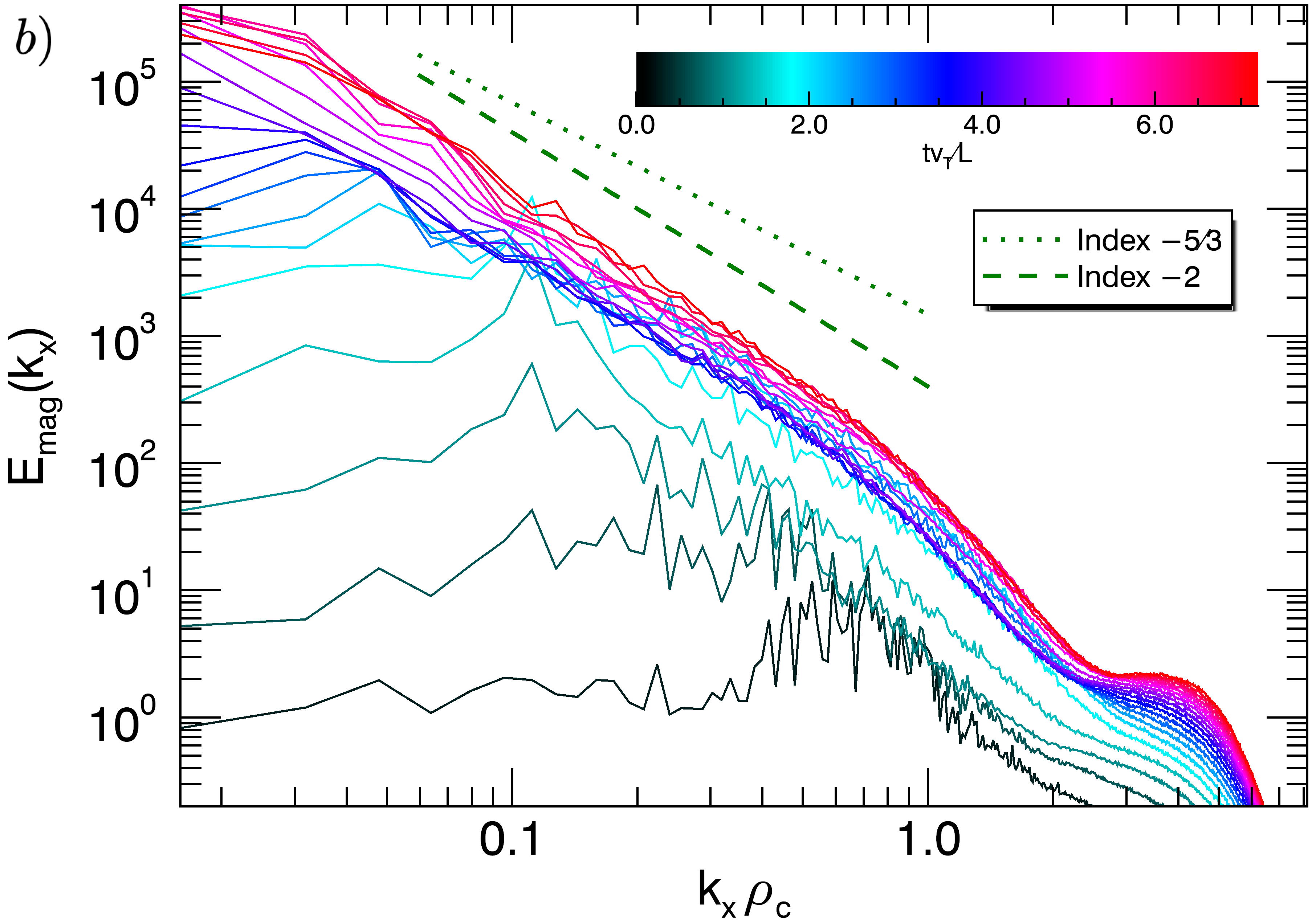}
   \centering
   \caption{\label{fig:energy} (a) Changes in the various energies (relative to initial magnetic energy); colors indicate magnetic (red), electric (blue), internal (magenta), flow kinetic (green), and negative of gravitational (black) energies. (b) Evolution of magnetic energy spectrum $E_{\rm mag}(k_x)$; power laws with indices $-5/3$ (dotted) and $-2$ (dashed) are shown for comparison.}
\end{figure}

The evolution of the overall energy is shown in Fig.~\ref{fig:energy}a. The RTI converts the free gravitational energy mainly into plasma internal energy and flow kinetic energy, in a multi-stage quasi-exponential manner. {The fastest growth of kinetic energy occurs during times when the plumes pierce through the cavity, tapping gravitational potential energy.} There is a minor conversion of free magnetic energy at early times ($t \lesssim 2 L/v_T$), but the magnetic energy experiences a net growth at late times ($t \gtrsim 5 L/v_T$) comparable to the plasma energy. At $t \gtrsim 6 L/v_T$, this accumulated magnetic energy is dissipated in the large-scale reconnection event. For reference, {in Fig.~\ref{fig:energy}} we show the {time evolution of the} magnetic energy spectrum {$E_{\rm mag}(k_x) = \int dy |\tilde{\boldsymbol{B}}(k_x,y)|^2$, where $\tilde{\boldsymbol{B}}(k_x,y)$ is the Fourier transform of the magnetic field vector in the $x$ direction at a given $y$ coordinate. This indicates} inverse transfer of energy to large scales until saturating with a broad power-law range, having an index varying between $-5/3$ and $-2$, reminiscent of MHD turbulence {\citep[e.g.,][]{goldreich_sridhar_1995, galtier_etal_2000}}.

\begin{figure}
\includegraphics[width=\columnwidth]{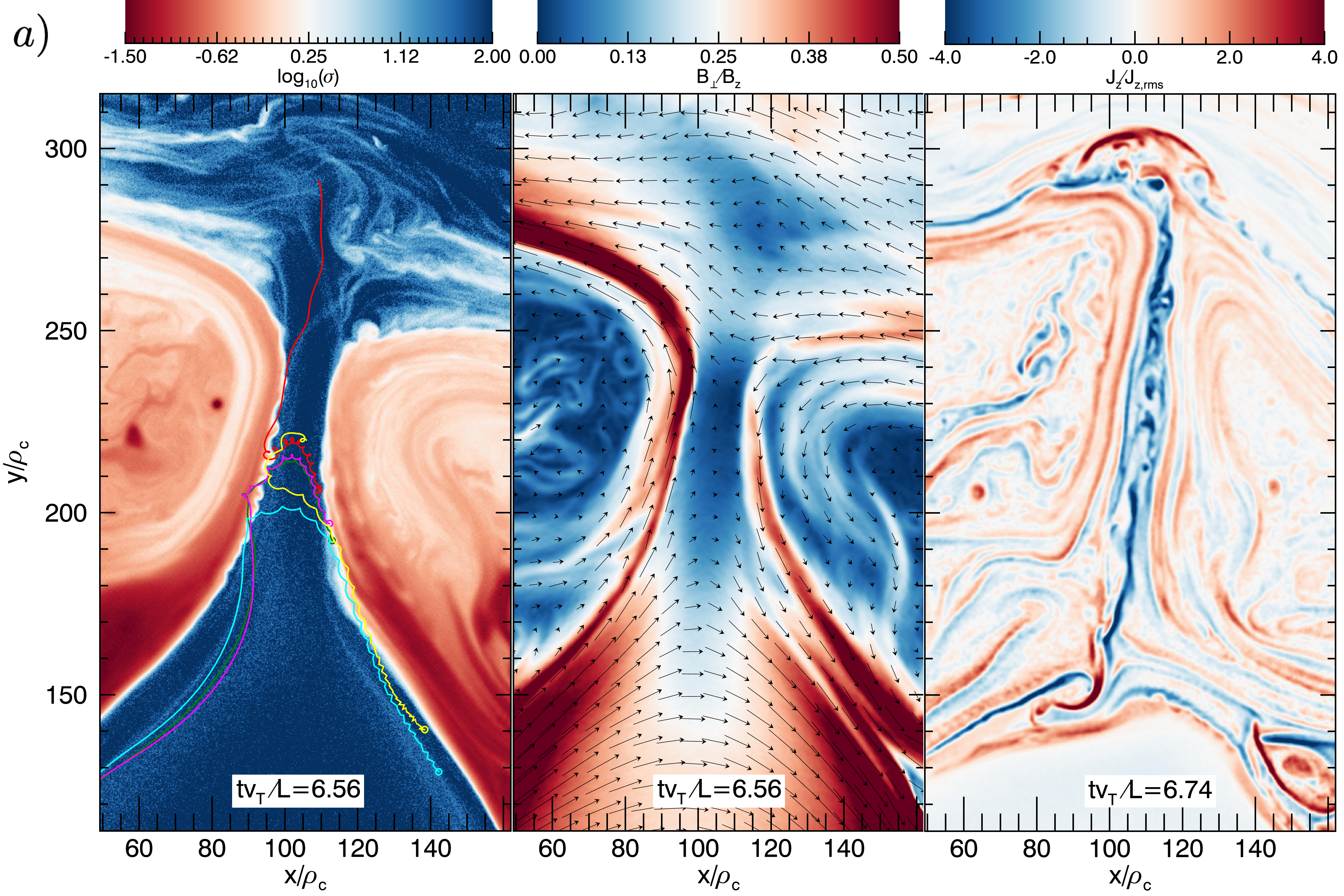}
\includegraphics[width=\columnwidth]{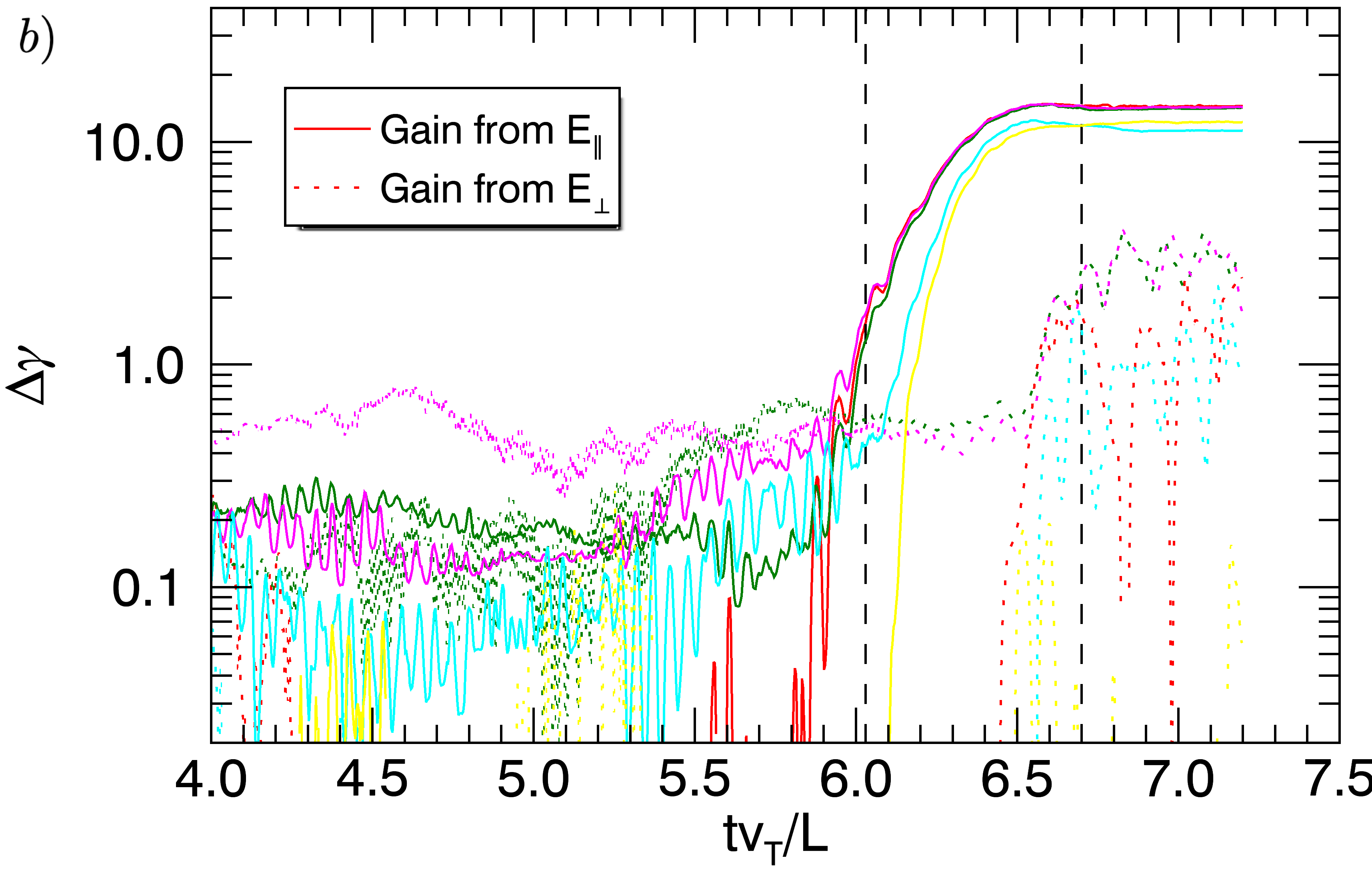}
   \centering
\caption{\label{fig:reconnection} (a) Zoom-in of the reconnection region from Fig.~\ref{fig:visual}b, showing magnetization $\sigma$ (with tracked particle trajectories during interval $6.0<tv_T/L <6.7$ overlaid, starting at the O symbols), ratio of in-plane to out-of-plane field $B_\perp/B_z$ (with magnetic field vectors overlaid), and out-of-plane current density $J_z$ (at a slightly later time, $tv_T/L = 6.75$). (b) Evolution of energy gain $\Delta\gamma$ from parallel (solid) and perpendicular (dotted) electric fields for same sample tracked particles (time interval $6.0 < tv_T/L < 6.7$ is marked by vertical dashed lines).}
\end{figure}

We now focus on the large-scale reconnection event at $t \approx 6.5 L/v_T$. This event initially processes the high-magnetization ($\sigma \gg 1$) cavity plasma, followed by the mixed ($\sigma \lesssim 1$) turbulent plasma in the lobes of the plume. The current sheet is displayed in Fig.~\ref{fig:reconnection}a, which shows $\sigma$, ratio of in-plane to out-of-plane field $B_\perp/B_z$ where $B_\perp = (B_x^2+B_y^2)^{1/2}$, and the out-of-plane current density $J_z$. The scenario resembles relativistic magnetic reconnection in the presence of a strong background field, $B_\perp/B_z \ll 1$, studied previously in local configurations \citep[e.g.,][]{werner_uzdensky_2017}. {However, the event differs from local studies in being driven by the expanding plume, and having a nonuniform (turbulent) upstream plasma.} We also overlay trajectories for 5 tracked particles that experience large energy gains; these tracked particles originate in the cavity and are co-located with the reconnection region as it forms. In Fig.~\ref{fig:reconnection}b, we show the tracked particle energy gain $\Delta \gamma(t) \equiv \gamma(t)-\gamma(0)$ from parallel and perpendicular electric fields. Here, $\gamma = (1 + p^2/m^2c^2)^{1/2}$ is the Lorentz factor for the particle with momentum $p$. There is strong, rapid energization by the parallel electric field ($E_\parallel = \boldsymbol{E} \cdot \boldsymbol{B}/B$) that occurs over a fraction of $L/v_T$ before the current sheet thins and tears. Subsequently, the energized particles interact with the turbulent outflows (visible in $J_z$) and gain additional energy from the (ideal) perpendicular electric field ($\boldsymbol{E}_\perp = \boldsymbol{E} - E_\parallel \boldsymbol{B}/B$). The highest energy particles mainly originate from the cavity, but there is also a significant fraction of high-energy particles that originate from the cold slab.

\begin{figure}
\includegraphics[width=0.95\columnwidth]{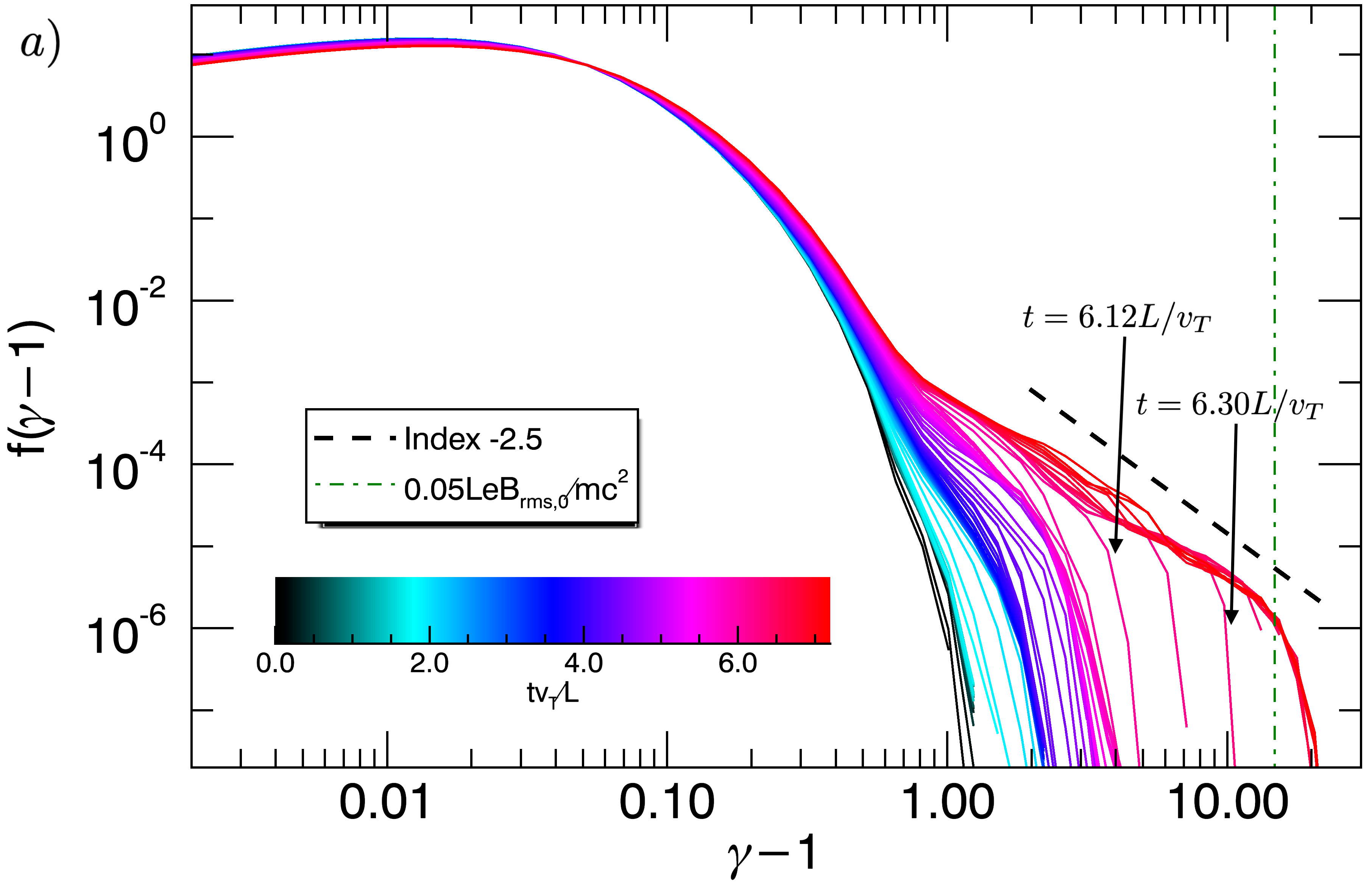}
\includegraphics[width=0.95\columnwidth]{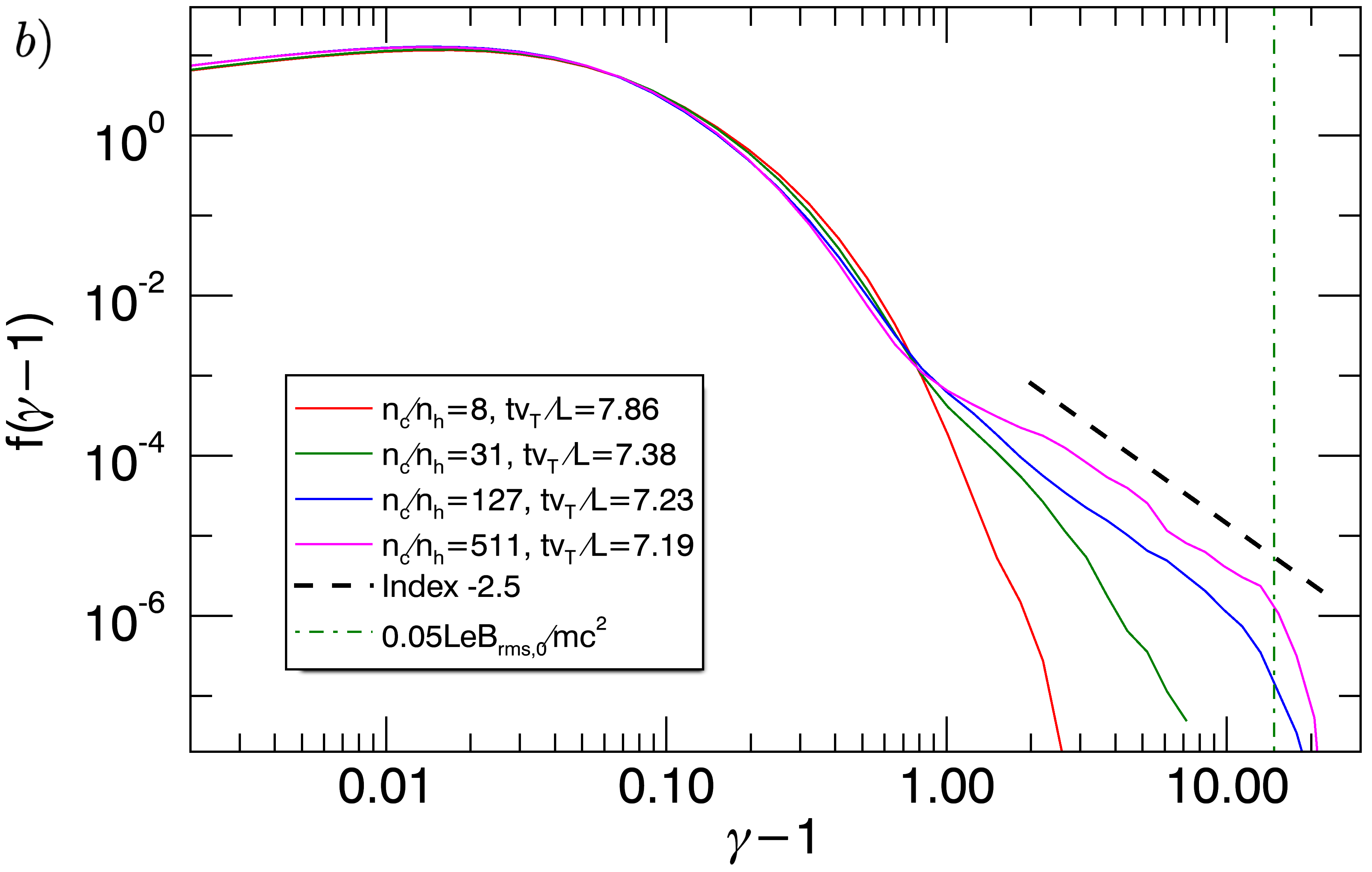}
\includegraphics[width=0.95\columnwidth]{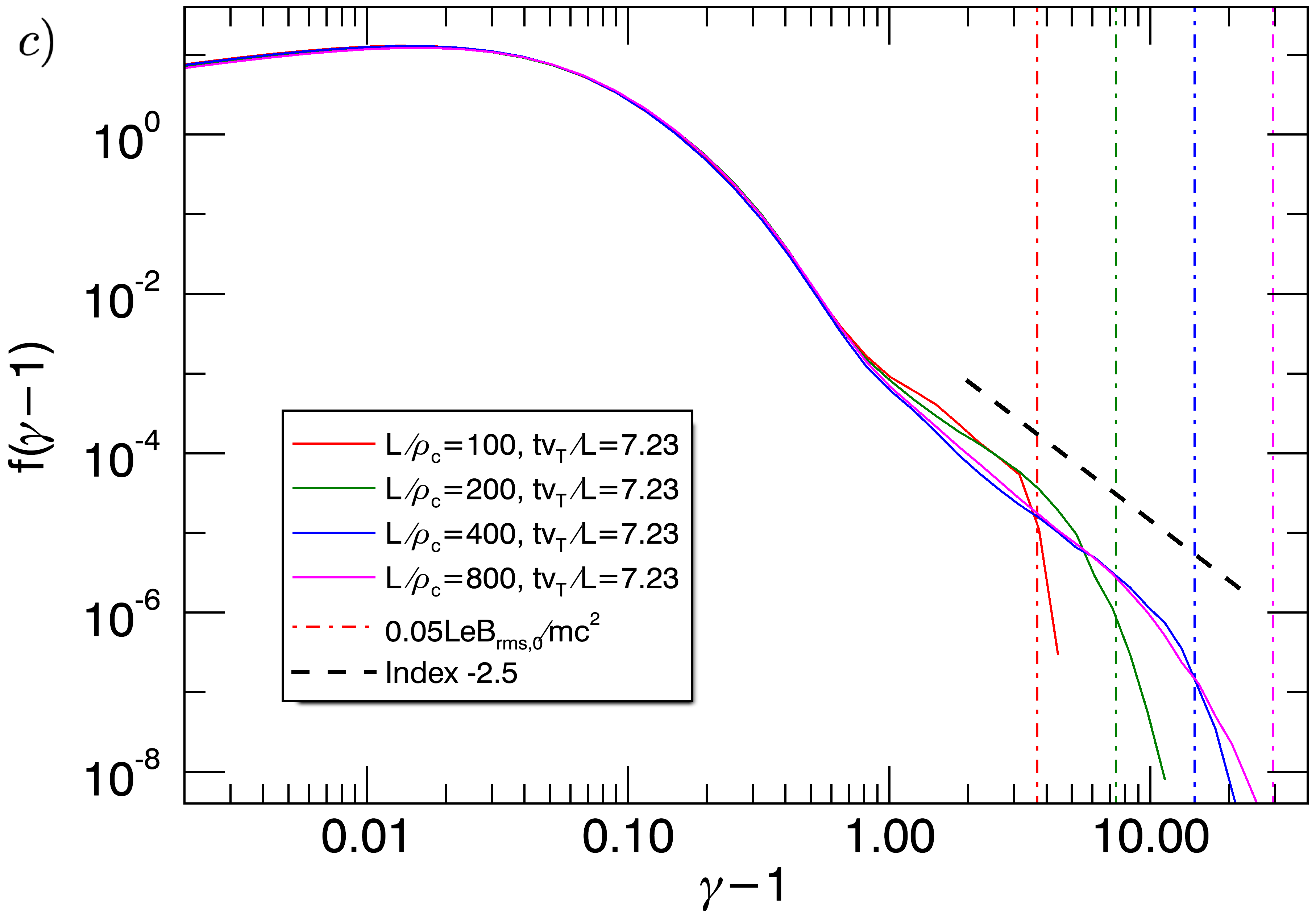}
   \centering
   \caption{\label{fig:dist}
   (a) Evolution of the particle kinetic energy distribution $f(\gamma-1)$; power-law $\propto (\gamma-1)^{-\alpha}$ with $\alpha=2.5$ is shown for reference (dashed) along with $\gamma_{\rm cut}-1 =0.05LeB_{{\rm rms},0}/mc^2$ (dash-dotted). (b) Distributions for $n_c/n_h \in \{ 8, 31, 127, 511 \}$ at fixed $L/\rho_c = 400$. (c) Distributions for~$L/\rho_c \in \{ 100, 200, 400, 800 \}$ at fixed $n_c/n_h = 127$.}
\end{figure}

In contrast to the laminar small-scale initial plume mergers, the large-scale reconnection event causes nonthermal particle acceleration that leads to an extended tail in the particle kinetic energy distribution $f(\gamma-1)$, as shown in Fig.~\ref{fig:dist}a. Prior to the reconnection event ($t \lesssim 6 L/v_T$), there is only a minor population of energetic nonthermal particles, resulting from interactions of the small-scale plumes. After the reconnection event ($t \gtrsim 6 L/v_T$), the tail rapidly broadens to a cutoff energy of $\gamma \sim 20$, and can be fit by a power law $f(\gamma-1) \sim (\gamma-1)^{-\alpha}$ with $\alpha \approx 2.5$. In this case, $\sim 0.1\%$ of all particles end up in the tail (with $\gamma > 1.8$).

Power-law tails of the distribution form only when $\sigma_h$ is sufficiently high, as shown from the $n_c/n_h$ parameter scan in Fig.~\ref{fig:dist}b, consistent with local models of magnetic reconnection \citep[e.g.,][]{guo_etal_2014, sironi_spitkovsky_2014, werner_etal_2016}. The power laws have $\alpha \lesssim 3$ when $n_c/n_h \gtrsim 100$ ($\sigma_h \gtrsim 15$). To demonstrate robustness of the results for varying domain sizes, in Fig.~\ref{fig:dist}c we show distributions from the system-size scan, having $L/\rho_c \in \{100, 200, 400, 800 \}$ with fixed $n_c/n_h = 127$. The high-energy cutoff of the distribution increases with $L/\rho_c$ for smaller sizes ($L/\rho_c < 400$), with an energy cutoff at $\gamma_{\rm cut}-1 \approx 0.1 (\gamma_{\rm max}-1)$ where $\gamma_{\rm max}-1 \equiv (1/2) L e B_{{\rm rms},0}/c$ is the maximum energy that can be confined by the domain given the initial rms magnetic field $B_{{\rm rms},0} = [(B_h^2+B_c^2)/2]^{1/2}$. At larger sizes ($L/\rho_c \gtrsim 400$), however, the distribution and its cutoff become independent of size. This trend is in qualitative agreement with local 2D relativistic reconnection simulations \citep[e.g.,][]{werner_etal_2016}, where a cutoff of $\gamma_{\rm cut} \sim 4 \sigma$ is reached on a fast timescale and further energization may occur on a longer timescale \cite{petropoulou_etal_2018, hakobyan_etal_2021}. The cutoff in our simulations is significantly below $4\sigma$, possibly due to the strong guide field at the reconnection sites. In 3D domains, the cutoff may be higher due to additional acceleration mechanisms \citep[e.g.,][]{zhang_sironi_giannios_2021}.

\section{Conclusions and discussion}

In this work, we demonstrated that the RTI is a viable mechanism for nonthermal particle acceleration in plasma regimes similar to black-hole accretion flows (moderately sub-relativistic ambient plasma on top of a magnetically dominated cavity with magnetic shear). The nonlinear development of the RTI leads to a rich evolution involving secondary instabilities, turbulence, and magnetic reconnection. The present numerical setup is idealized in being local, {and as a consequence, the late-time evolution may be influenced by the boundaries of the domain. Nevertheless,} we expect the process of RTI plumes merging into a large-scale plume {(with size comparable to the domain or scale height)} that relaxes via magnetic reconnection (which ultimately accelerates the particles) to be generic and robust.

The high-energy particle distribution arising from RTI-induced magnetic reconnection, $f(\gamma-1) \sim (\gamma-1)^{-\alpha}$ with $\alpha \approx 2.5$, meets the requirements for synchrotron-radiating electrons needed to explain Sgr A* NIR flares. Specifically, the spectral luminosity $\nu L_\nu \sim \nu^{b}$ with indices $0 \lesssim b \lesssim 0.5$ is measured during high-flux states, implying $\alpha = 3 - 2 b$ in the range $3 \gtrsim \alpha \gtrsim 2$ \citep{dodds-eden_etal_2009, abuter_etal_2021, ponti_etal_2017, boyce_etal_2022}. We next argue that when accounting for realistic electron/positron temperatures, the number of accelerated particles is also sufficient to supply NIR flares with observed luminosities up to $\mathcal{P}_{\rm tot} \sim 10^{35}$ erg/s \citep{yusef-zadeh_etal_2009, dodds-eden_etal_2009, abuter_etal_2018, ponti_etal_2017, fazio_etal_2018, abuter_etal_2021, boyce_etal_2022}.

For an electron (or positron) emitting in the NIR range at $\varepsilon\sim 0.7$ eV, the typical Lorentz factor is $\gamma_{\rm NIR} = \sqrt{\varepsilon m_{\rm e} c/(e\hbar B)}\sim 1.4 \times 10^3$, for a $B = 30$ G magnetic field at the emitting radius $\sim 10 r_{\rm g}$ \citep{abuter_etal_2018, abuter_etal_2021, akiyama_etal_2022}, where $r_{\rm g} = G M/c^2 = 6.1 \times 10^{11}$ cm is the gravitational radius for Sgr A* with black-hole mass $M = 4.1 \times 10^6 M_{\odot}$ in terms of solar mass. The total number of NIR synchrotron-radiating electrons and positrons in the flaring emission region is approximately $N_{\rm NIR} = 8\pi \mathcal{P}_{\rm tot} / (4 \sigma_{\rm T} c \gamma_{\rm NIR}^2 B^2/3)$, where $\sigma_{\rm T} = 8 \pi e^4/ (3 m_{\rm e}^2 c^4)$ is the Thomson cross section. This implies a corresponding number density $n_{\rm NIR} = N_{\rm NIR} / (\pi R^2 H) = 6\mathcal{P}_{\rm tot} / (\sigma_{\rm T} c \gamma_{\rm NIR}^2 B^2 R^2 H) \sim 50$ cm$^{-3}$, where $R$ and $H$ are the radius and height of the emitting region (assumed cylindrical), for which we took fiducial values of $R = 5 r_{\rm g}$ and $H = 60 r_{\rm g}$, as inferred from the MHD simulation described in this work. The density of NIR-radiating particles is related to the average density in the power law by
\begin{align}
\frac{n_{\rm NIR}}{n_{\rm PL}} \sim \frac{(\gamma_{\rm NIR}-1) f(\gamma_{\rm NIR}-1)}{(\gamma_{\rm min}-1) f(\gamma_{\rm min}-1)} \sim \left(\frac{\gamma_{\rm NIR}-1}{\gamma_{\rm min}-1}\right)^{1-\alpha} \, ,
\end{align}
where $\gamma_{\rm min}$ is the Lorentz factor at which particles are injected into the power law tail. As indicated by our PIC simulations, the injected electrons may come from either the cavity or ambient flow. In both cases, we expect $\gamma_{\rm min} \sim 100$, noting that ambient average electron Lorentz factor is $\gamma \approx 3 T/m_e c^2 \sim 30$ for Sgr A* \cite{akiyama_etal_2022}. For $\alpha = 2.5$, we then find $n_{\rm NIR}/n_{\rm PL} \sim 2\times10^{-2}$. Based on quiescent submillimeter emission, the electron density in the ambient accretion flow is $n_{\rm e, amb} \approx 10^{6}$ cm$^{-3}$ \citep{akiyama_etal_2022}. Therefore, to supply the flares, the ratio of accelerated particles to ambient electrons must be $n_{\rm PL}/n_{\rm e, amb} = (n_{\rm PL}/n_{\rm NIR}) (n_{\rm NIR}/n_{\rm e, amb}) \sim 2.5\times 10^{-3}$. Our PIC simulations indicate that $n_{\rm PL}/n_{\rm e,amb} \sim 10^{-3}$ for $\sigma_h = 80$, which is within a factor of a few of the required supply. We expect that higher $\sigma_h$ would increase $n_{\rm PL}/n_{\rm e,amb}$ and harden the distribution ($\alpha \to 2$), to a degree that can easily supply the NIR flares.

Finally, we note that the synchrotron radiative cooling time at energies near $\gamma_{\rm NIR}$ may be comparable to the typical flare duration in Sgr A*.  Thus, the flare evolution may be shaped by radiative cooling at energies near and above $\gamma_{\rm NIR}$. However, the particle acceleration via magnetic reconnection will happen on a fraction of a dynamical time, and so will be rapid with respect to the cooling. Radiative cooling may limit the extent of the power-law distribution at energies above the near-infrared value $\gamma_{\rm NIR}$.

In a 3D domain, additional degrees of freedom may enhance the conversion of free energy, increasing the acceleration efficiency \citep{zhang_sironi_giannios_2021} as long as there is sufficient inverse energy transfer to large-scale magnetic fields. {The hydrodynamic RTI is known to have enhanced inverse energy transfer in 2D when compared to 3D \citep{zhao_etal_2022}; in MHD, however, there is expected to be a forward energy cascade and simultaneous inverse magnetic energy transfer in both 2D and 3D \citep{zrake_2014, brandenburg_etal_2015, zhou_etal_2020, hosking_schekochihin_2021}.} Also, flaring signatures may be enhanced if the cavity is filled with pre-accelerated relativistic particles (expected from the reconnection event that produces the cavity \citep{hakobyan_etal_2023,galishnikova_etal_2023}). Future work should thus consider 3D simulations with electron-ion plasma in the dense slab and ultra-relativistic, radiating pair plasma in the cavity.

\begin{acknowledgments}

The authors thank Lorenzo Sironi, Jim Stone, Greg Werner, Koushik Chatterjee, Daryl Haggard, Gunther Witzel, Sebastiano von Fellenberg, and Eliot Quataert for useful conversations{, and the anonymous referees for constructive feedback.} Research at the Flatiron Institute is supported by the Simons Foundation. This work used the Extreme Science and Engineering Discovery Environment (XSEDE), which is supported by National Science Foundation grant number ACI-1548562. This work used the XSEDE supercomputer Stampede2 at the Texas Advanced Computer Center (TACC) through allocation TG-PHY160032 \citep{xsede}. Support for this work was provided by NASA through the NASA Hubble Fellowship grant HST-HF2-51518.001-A awarded by the Space Telescope Science Institute, which is operated by the Association of Universities for Research in Astronomy, Incorporated, under NASA contract NAS5- 26555. {A.P. acknowledges support by the NASA grant 80NSSC22K1054.}

\end{acknowledgments}

\appendix

\section{Effect of rotation angle} \label{sec:rot}

\begin{figure*} 
\includegraphics[width=0.65\columnwidth]{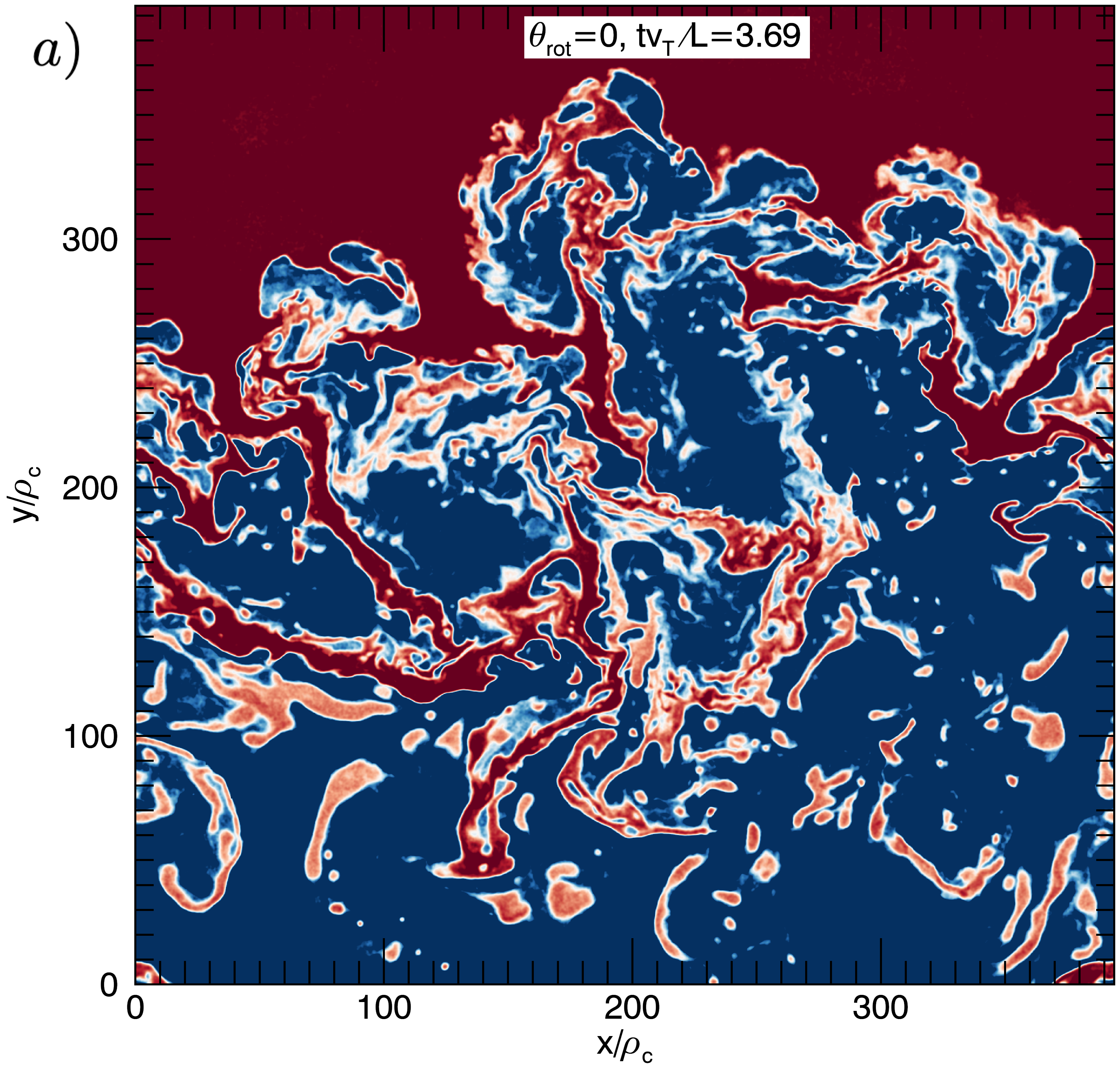}
\includegraphics[width=0.65\columnwidth]{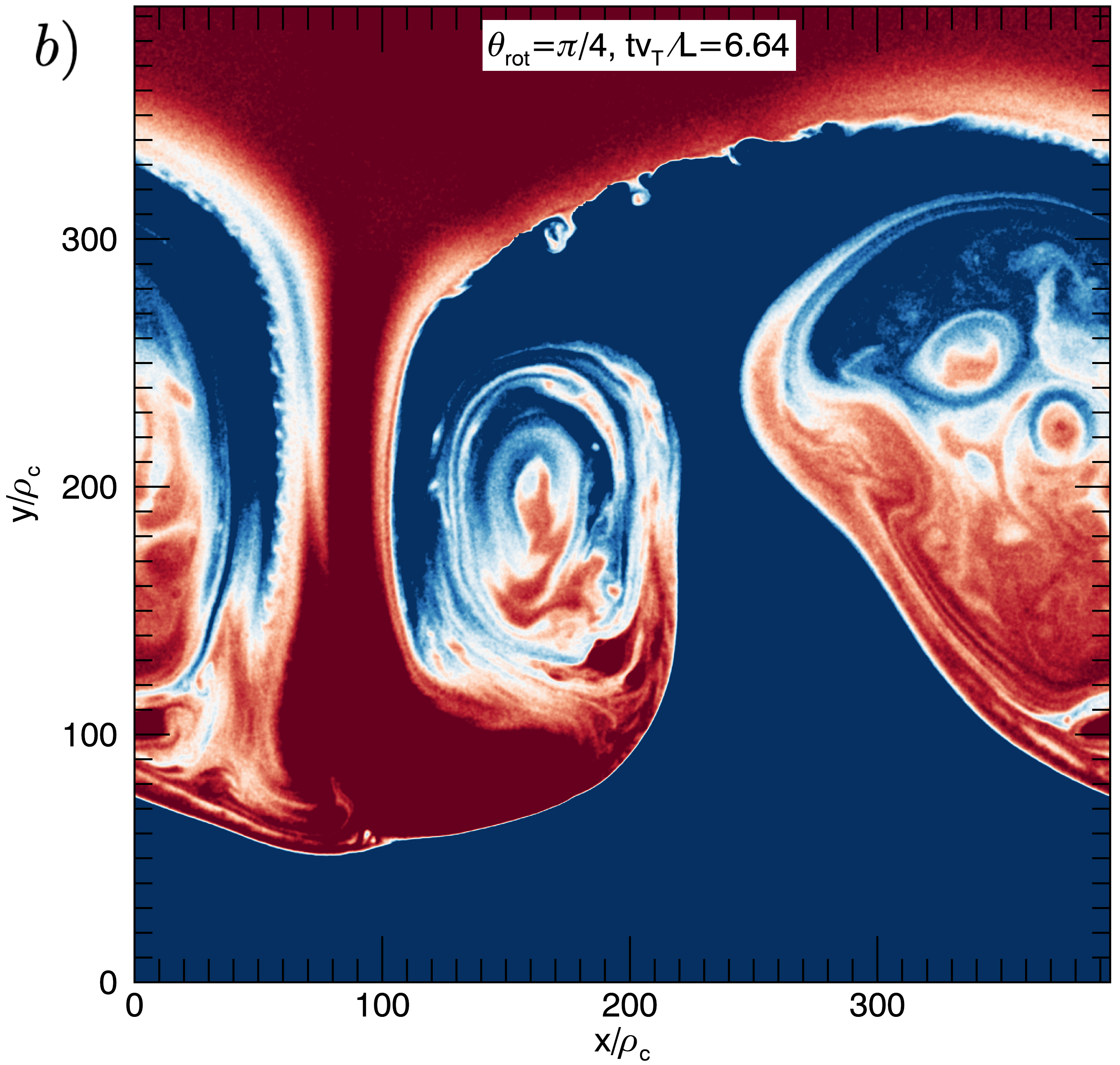}
\includegraphics[width=0.65\columnwidth]{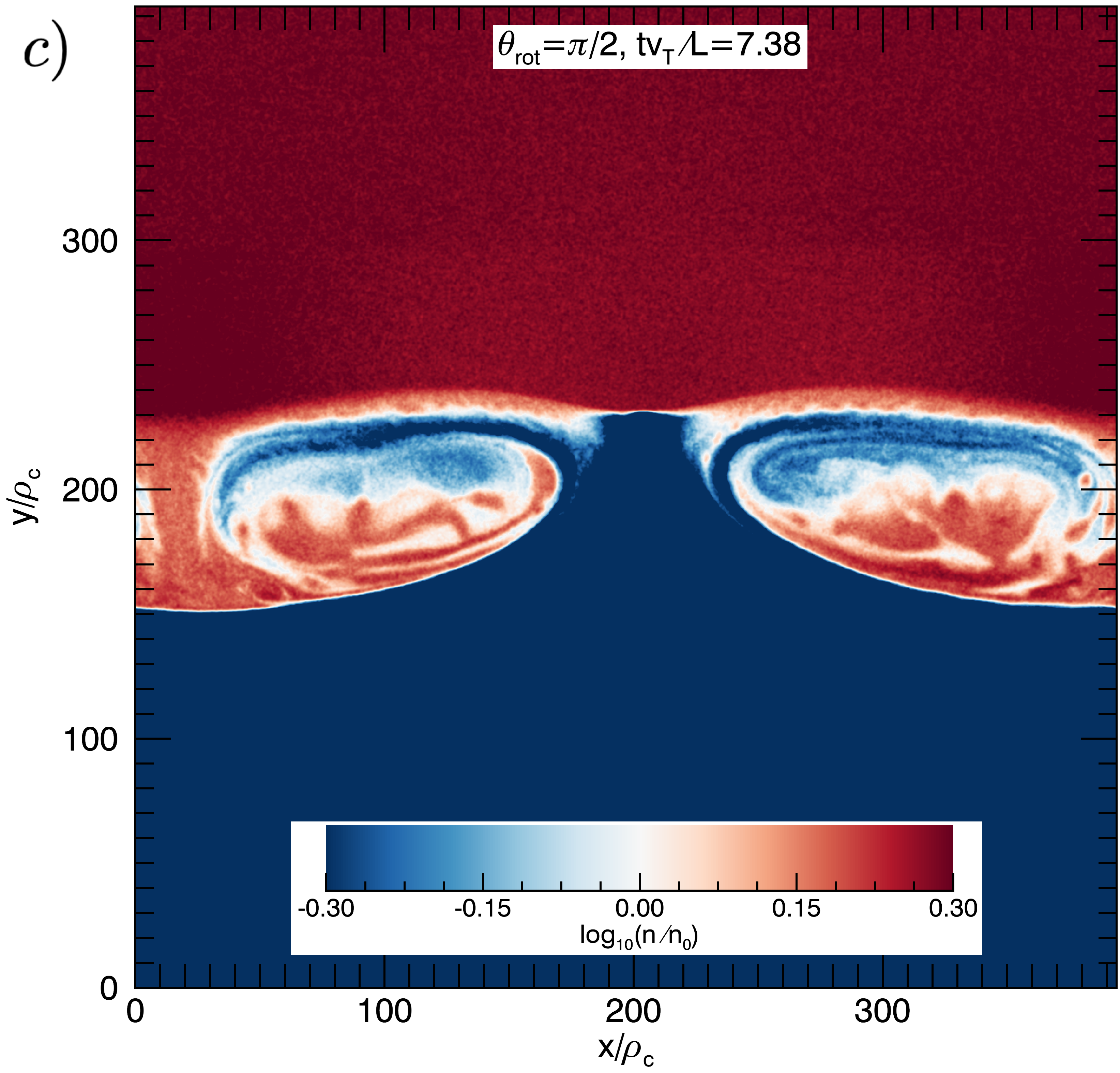}
   \centering
   \caption{\label{fig:thetarot}
Images of the particle number density for simulations with varying magnetic rotation angle: (a) $\theta_{\rm rot} = 0$, (b) $\theta_{\rm rot} = \pi/4$, and (c) $\theta_{\rm rot} = \pi/2$. This set of simulations has $n_c/n_h = 31$. Times are chosen as indicated to show the late nonlinear development of the instability.}
\end{figure*}

\begin{figure} 
\includegraphics[width=0.95\columnwidth]{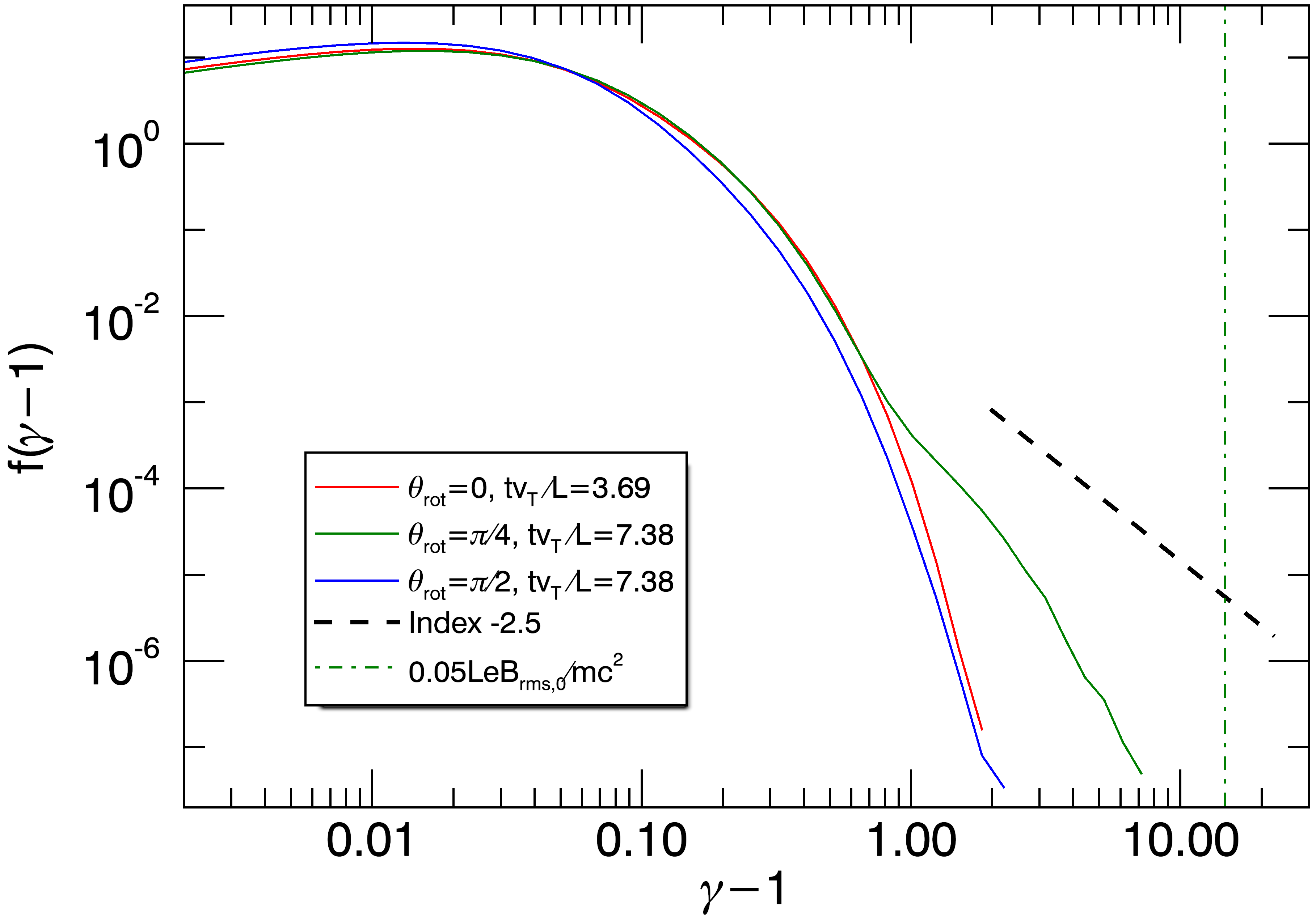}
   \centering
   \caption{\label{fig:distrot}
Particle kinetic energy distribution $f(\gamma-1)$ for simulations shown in Fig.~\ref{fig:thetarot}: $\theta_{\rm rot} = 0$ (red), $\theta_{\rm rot} = \pi/4$ (green), and $\theta_{\rm rot} = \pi/2$ (blue). Power-law~$\propto (\gamma-1)^{-\alpha}$ with $\alpha=2.5$ shown for reference (dashed) along with $\gamma_{\rm cut}-1 =0.05LeB_{{\rm rms},0}/mc^2$ (dash-dotted). An earlier time is chosen in the $\theta_{\rm rot} = 0$ case due to eventual transport through the boundaries.}
\end{figure}

Since magnetic shear is a key ingredient in our problem, we now describe the effect of varying the magnetic rotation angle $\theta_{\rm rot}$. In Fig.~\ref{fig:thetarot}, we show late-time density images for a set of simulations with $n_c/n_h = 31$ and varying rotation angle: $\theta_{\rm rot} \in \{ 0, \pi/4, \pi/2 \}$. In Fig.~\ref{fig:distrot}, we show particle energy distributions for the same simulations.

For the case of no field rotation, $\theta_{\rm rot} = 0$, the magnetic field is directed entirely out of the 2D domain. As there is no magnetic tension, the two regions freely mix, as shown in Fig.~\ref{fig:thetarot}a. The RTI grows at small scales and does not exhibit significant transfer of energy to large-scale plumes. Due to the orientation of the fields, there is no magnetic reconnection. As a consequence, free gravitational energy is mainly converted to bulk kinetic energy, and there is no particle acceleration.

As described in the paper, the fiducial case of $\theta_{\rm rot} = \pi/4$ develops small-scale plumes that merge to eventually form a large-scale plume, as shown in Fig.~\ref{fig:thetarot}b. Due to amplification of the magnetic fields, a significant amount of free magnetic energy develops that is eventually released by magnetic reconnection. This case has the necessary complexity and self-organization for efficient particle acceleration.

For the case with extreme field rotation, $\theta_{\rm rot} = \pi/2$, we find that plasmoid-unstable, asymmetric magnetic reconnection occurs at the interface on a faster timescale than the growth of the RTI fingers. Reconnection mixes the plasma along the interface, which prevents the growth of the fingers and thus inhibits mixing of the plasma between the two regions. The plasmoids merge and grow in size until stalling in the state shown in Fig.~\ref{fig:thetarot}c. Since magnetic reconnection at the interface is asymmetric, with one side having a high plasma beta ($\beta_c = 4$), it is ineffective at accelerating particles \citep{mbarek_etal_2022}.

 The case with extreme magnetic shear appears to be the most relevant for the cavity interface in the global MHD simulation described in this work. However, a broader shear layer or other physical effects may need to be needed to stabilize the interface to magnetic reconnection.


%

\end{document}